# Laser Stimulated Grain Growth in 304 Stainless Steel Anodes for Reduced Hydrogen Outgassing.

D. Gortat[a*], M. Sparkes[a], S.B. Fairchild[b], P.T. Murray[c],
M.M. Cahay[d], T.C. Back[c], G.J. Gruen[c], N.P. Lockwood[e] and W. O'Neill[a].

a- Institute for Manufacturing, University of Cambridge, 17 Charles Babbage Road, Cambridge CB3 0FS, UK. (dg458@cam.ac.uk).
b- Materials and Manufacturing Directorate, Air Force Research Laboratory, WPAFB, OH 45433, USA.
c- Research Institute, University of Dayton, Dayton, OH 45469, USA.
d- Spintronics and Vacuum Nanoelectronics Laboratory, University of Cincinnati, Cincinnati, OH 45221, USA.
e- Directed Energy Directorate, Air Force Research Laboratory, Kirtland, AFB, NM 87117, USA.
*- Corresponding author.

## ABSTRACT

Metal anodes in high power source (HPS) devices erode during operation due to hydrogen outgassing and plasma formation; both of which are thermally driven phenomena generated by the electron beam impacting the anode's surface. This limits the lowest achievable pressure in an HPS device, which reduces its efficiency. Laser surface melting the 304 stainless steel anodes by a continuous wave fiber laser showed a reduction in hydrogen outgassing by a factor of ~4 under 50 keV electron bombardment, compared to that from untreated stainless steel. This is attributed to an increase in the grain size (from 40 - 3516 μm$^2$), which effectively reduces the number of characterized grain boundaries that serve as hydrogen trapping sites, making such laser treated metals excellent candidates for use in vacuum electronics.

## 1. INTRODUCTION

### 1.1 Anodes in High Power Source Devices

High power source (HPS) devices generate electromagnetic radiation typically spanning frequencies of 1 GHz to 100 GHz[1]. Their applications include microwave generation[1,2], vacuum electronics[2–4] and particle acceleration[5,6]. HPSs operate as vacuum tubes where the interaction of a stream of electrons with a magnetic field generates microwaves. Electrons are produced by a negatively charged, heated component, the cathode, and accelerated by a positively charged component, the anode, to produce a high velocity electron stream[2]. The conversion of the kinetic energy of the electron current to coherent electromagnetic radiation experiences no energy or efficiency loss and no internal heat generation as a result of the electron stream moving without collision through the evacuated region[7]. However, energetic electrons incident on a conducting surface cause extensive heating, especially on anode surfaces which are constantly bombarded with electrons resulting in a pressure rise inside any microwave power electronics[7]. The pressure rise in HPSs during their operation is attributed to outgassing, predominantly of the hydrogen atoms stored in anode's bulk due to their prior exposure to humid air[2,7,8]. Plasma may also form at the surface of the anode due to near surface ionization of the outgassed neutral atoms by the anode's desorbed electrons or secondary electrons. This can cause pulse-shortening of an HPS device[9] and, in the long-term, erosion of the anode[10]. This paper presents a hydrogen outgassing reduction technique at the core of which is a microstructural transformation caused by laser melting the surface of the HPS metal anodes.

### 1.2 The Mechanism of Anode's Hydrogen Outgassing

Hydrogen exists as atomic hydrogen at defects such as grain boundaries (GBs) in the anode's bulk[11]. The binding energy of a GB is larger than the activation energy of hydrogen diffusion[11] and, therefore, the hydrogen atom is trapped within the vicinity of a GB. GB's binding energy is, however, temperature-dependent (it decreases with temperature rise[12]), increasing the temperature releases hydrogen and increases the outgassing rate of the sample. In polycrystalline metals, at high temperatures, the hydrogen diffusion along the GBs is mainly due to high energy boundaries ("Random", $\Sigma > 29$). The $\Sigma$ value is the reciprocal of the fraction of lattice points in the boundaries that coincide between the two adjoining grains[13]. It is calculated within this work from the more restrictive Palumbo-Aust criteria ($\theta \leq 15 \Sigma^{-5/6}$)[14], where $\theta$ is the maximum deviation angle from the computed misorientation angle of the GB[15,16]. Special boundaries or low energy boundaries (between $\Sigma 3$ and $\Sigma 29$) will contain less hydrogen atoms due to their low binding energy levels[17–22] and consequently reduce the overall sample's diffusion. Metals with low hydrogen outgassing rates, such as austenitic stainless steel, are the most common materials for vacuum applications, due to low hydrogen outgassing properties[8]. To further reduce outgassing in such metals, several treatments have been proven effective: baking[23–27], vacuum baking[24,26,28,29], polishing[29–32] and surface treatments to create oxide or other protective surface films (e.g. titanium nitride or boron nitride[33]). We present a laser surface melting (LSM) technique which reduces hydrogen outgassing of stainless steel by stimulating the grain growth at the surface of the substrate with no post-processing to preserve the benefits of the treatment.

## 2. EXPERIMENTAL

SPI™ G3 Yb non-polarized fiber laser, wavelength 1.064 μm, an output beam $M^2 = 2$ and maximum output power of 20 W in continuous wave (CW) mode was used for melting the surface of 3 mm and 0.6 mm thick 304 SS plates. The lens used was a Jenoptic™ 03-90FT fused silica lens, with focal length of 125 mm. The samples were processed at room temperature with a continuous flow of nitrogen supplied into the capped stage to minimize the sample's oxidation. Greisenger™ GOX 100T oxygen meter was used to track the oxygen levels to 0.1 – 0.2 % in the capped stage. Ethanol was used to clean/remove contaminants from the sample's surface prior to laser processing. For characterization, an Olympus BX51 optical microscope with Jenoptic™ ProgRes C10+ CCD camera was used to record optical images of the laser penetration depth, where chemical etching was used to contrast the laser affected area. Chemical composition of the etchants was 10g $FeCl_3$, 30 ml HCl, 120 ml water. For characterizing the grain boundaries an FEI/Philips™ XL30 scanning electron microscope (SEM) with electron backscatter diffraction (EBSD) analysis capability was used. Visual representation of the grains was generated by the HKLTango™ software in a form of combined three angle Euler maps for a three-dimensional representation of the samples' crystal lattices. The hydrogen outgassing testing was done with the help of Anode Materials Characterization System (AMCS), a high vacuum system where a beam of electrons impacts the tested anode at normal incidence. A 50 keV electron beam 1.6 mm in diameter at 60 second intervals was used. A SRS™ RGA100 residual gas analyzer detected the outgassed elements from the samples. The electron current density at the sample surface was approximately 16.4 mA/cm$^2$ at base pressure of $5\times10^{-10}$ Torr.

# 3. RESULTS AND DISCUSSION

The 304 SS was processed by raster scanning the CW laser beam across the surface in a uniform pattern, Table 1.

Table 1. Laser parameters.

| Scanning velocity (mm/s) | 2.25 |
|---|---|
| Output power (W) | 12, 17 |
| Spot size diameter (μm) | 39.4 |
| Hatch spacing (mm) | 0.03 |

Figure 1 shows a longitudinal cross-section view of the of the SS treated samples 3 mm and 0.6 mm thick, where the treated region is comprised of a laser surface melted zone (LSMZ) and heat-affected zone (HAZ). The depth of the LSMZ for 3 mm sample treated with 13.54 kJ/cm$^2$ laser energy density is approximately 9.7 μm, for 0.6 mm sample treated with 19.17 kJ/cm$^2$ the LSMZ is ~11 μm.

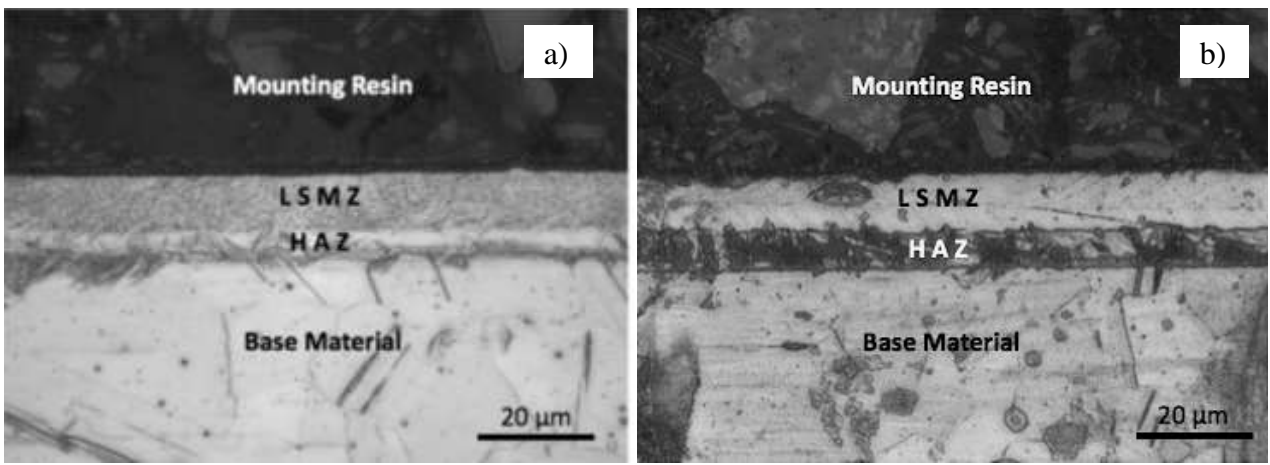

Figure 1. Optical image of the section view of the raster scanned 304 SS samples, a) 3 mm, 13.54 kJ/cm$^2$, b) 0.6 mm, 19.17 kJ/cm$^2$ laser energy density.

Shown in Figure 2(a) and Figure 2(b) are three angle Euler maps of the untreated and treated samples, demonstrating the microstructural changes induced by 13.54 kJ/cm$^2$ laser energy density inside the LSMZ of the 3 mm SS sample. It can be observed that the grains in the laser treated sample, Figure 2(b), are elongated in the direction of the laser scan and have increased in size; per 0.12 mm$^2$ area of the sample's surface the number of grains was reduced from 1020 to 617 with some grains growing from 40 to 3516 μm$^2$ indicating abnormal grain growth, Table 2.

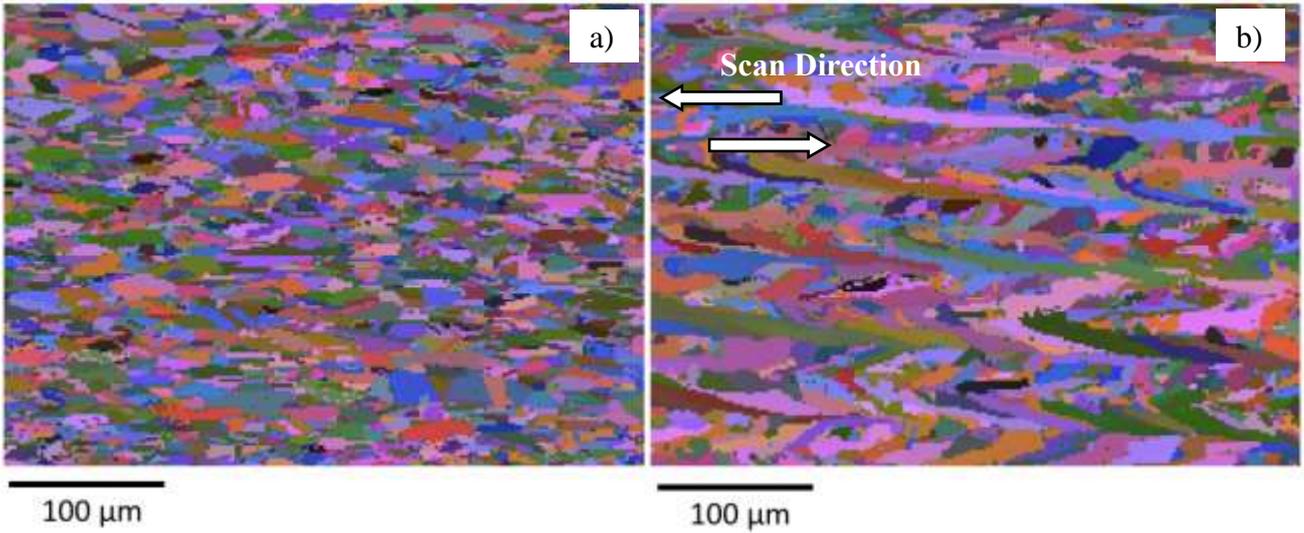

Figure 2. Combined three angle Euler maps of (a) untreated and (b) laser treated 304 SS, 3 mm, 13.57 kJ/cm². Scan step size 2 μm.

The 0.6 mm SS sample is noted to have smaller average grain size compared to that of 3 mm SS sample, Table 2. The treatment of 0.6 mm sample with laser energy density of 19.17 kJ/cm² showed further increase in grain elongation and size, Figure 3(b). The number of grains was reduced from 1083 to 96 per 0.17 mm² surface area. Some grains reached 12040 μm² in area also indicating the abnormal grain growth, Table 2.

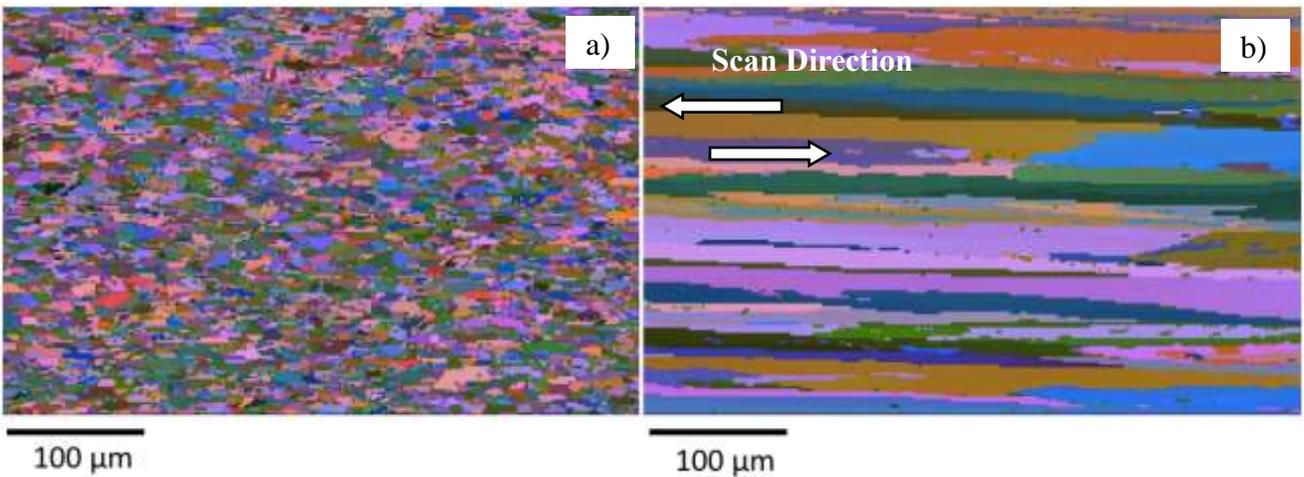

Figure 3. Combined three angle Euler maps of (a) untreated and (b) laser treated 304 SS, 0.6 mm, 19.17 kJ/cm². Scan step size 2 μm.

Table 2. Grain measurements showing average grain size and range for untreated and treated 304 SS.

| Grain area | 3 mm sample, 0.12 mm² surface area | | 0.6 mm sample, 0.17 mm² surface area | |
|---|---|---|---|---|
| | Untreated | Treated | Untreated | Treated |
| Average (μm²) | 99.88 | 201.59 | 73.35 | 1756.17 |
| Minimum (μm²) | 40.00 | 40.00 | 40.00 | 40.00 |
| Maximum (μm²) | 628.00 | 3516.00 | 464.00 | 12040.00 |

We attribute the abnormal grain growth to the laser scanning strategy. Grain nucleation appears to have little dependency on original grain size, regulated only by laser power, spot size, speed and hatch spacing. Larger laser energy density prompted higher melting temperatures per same surface area of the sample, causing the grains to merge and grow at the expense of the neighboring grains. In terms of types of grains and GBs obtained, both treated samples exhibit a reduction in the fraction of GBs, Table 3.

Table 3. Grain boundary classification for untreated and treated 304 SS.

| Sigma value | 3 mm sample, 0.12 mm² surface area | | 0.6 mm sample, 0.17 mm² surface area | |
|---|---|---|---|---|
| | Untreated | Treated | Untreated | Treated |
| $\Sigma\ 3$ (%) | 33.81 | 15.39 | 30.40 | 9.49 |
| $3 \leq \Sigma \leq 29$ (%) | 35.11 | 16.66 | 32.00 | 10.31 |
| $29 < \Sigma \leq 49$ (%) | 0.07 | 0.05 | 0.08 | 0.05 |
| Total $\Sigma$ (3-49) (%) | 35.19 | 16.71 | 32.08 | 10.37 |

The measured GB reduction is largely in the dominant $\Sigma\ 3$ boundaries. The total number of measured GBs was reduced by 18.47 % per 0.12 mm² for 3 mm and 21.72 % per 0.17 mm² for 0.6 mm SS samples, which indicates the benefit of increased laser energy density for LSM grain boundary reduction. Figure 4 shows hydrogen outgassing results of the 3 mm SS sample.

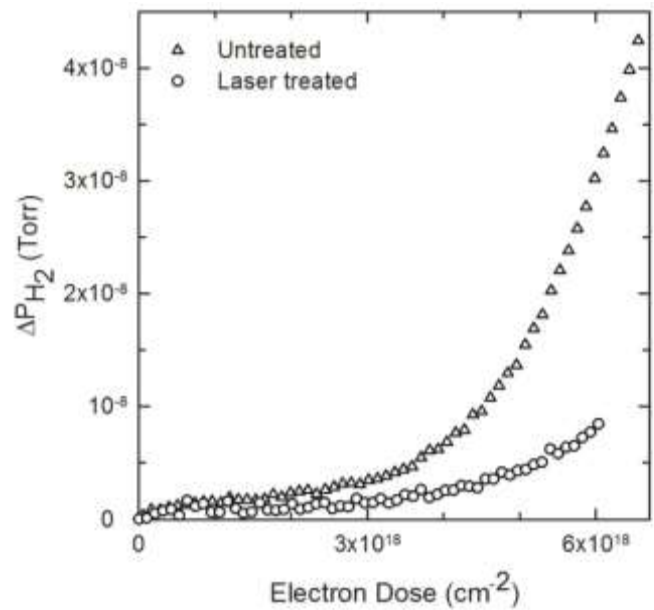

Figure 4. Outgassing results showing the change in $H_2$ partial pressure with electron dose during 60 second 3 mm 304 SS electron irradiation.

The results are presented as the change in hydrogen atom partial pressure above baseline with electron dose. An incubation period is observed between the electron dose of 0 to ~ $3 \times 10^{18}$ cm⁻². In this period the atomic hydrogen diffusion is impeded due to sample's heating until its recombination at the surface of the sample into hydrogen molecules for surface desorption[34]. The laser treated sample exhibits a smaller slope, suggesting a lower rate of diffusion of atomic hydrogen to the surface because of the decreased number of grain boundaries. From the GB characterization results obtained for 0.6 mm SS sample, it is suspected that the hydrogen outgassing testing of 304 SS treated with 19.17 kJ/cm$^2$ or higher average laser energy density should show further hydrogen outgassing reduction.

## 4. CONCLUSION

We demonstrate that laser melting the surface of 304 stainless steel reduces its hydrogen diffusion via grain growth. The measured grain size in 3 mm SS sample per 0.12 mm$^2$ surface area counted 40 - 3516 μm$^2$ maximum increase, for 0.6 mm SS sample 40 – 12040 μm$^2$ per 0.17 mm$^2$. The 3 mm SS sample was bombarded with 50 keV electron beam and showed a factor of ~4 hydrogen outgassing reduction compared to untreated 3 mm SS sample. This is attributed to the decrease by 18.47 % in the measured grain boundary count.